\documentclass[reprint, amsmath, amssymb, aps, prd,showpacs, superscriptaddress,
]{revtex4-1}

\usepackage{graphicx}% Include figure files
\usepackage{dcolumn}% Align table columns on decimal point
\usepackage{bm}% bold math
\usepackage{mathtools}
\usepackage[utf8]{inputenc} 
\usepackage{xcolor}
\usepackage{lineno}
\usepackage{lipsum}
\usepackage{aas_macros}
\usepackage{url}
\usepackage[colorlinks]{hyperref}
\usepackage{booktabs}
\usepackage{physics}
\usepackage{xspace}
\usepackage{multirow}

\AtBeginDocument{
	\heavyrulewidth=.08em
	\lightrulewidth=.05em
	\cmidrulewidth=.03em
	\belowrulesep=.65ex
	\belowbottomsep=0pt
	\aboverulesep=.4ex
	\abovetopsep=0pt
	\cmidrulesep=\doublerulesep
	\cmidrulekern=.5em
	\defaultaddspace=.5em
}
\definecolor{dodgerblue}{HTML}{1E90FF}
\definecolor{magmaorange}{HTML}{EA8246}
\definecolor{viennared}{HTML}{DA0A14}
\definecolor{ctorange}{HTML}{FF6C0C}
\definecolor{wales}{HTML}{ff0038}
\definecolor{benettongreen}{HTML}{009421}
\definecolor{ferrarired}{HTML}{ff2800}
\definecolor{austriawienpurple}{HTML}{441678}

\hypersetup{
     colorlinks=true,
     linkcolor= dodgerblue,
     filecolor= dodgerblue,
     citecolor = dodgerblue,      
     urlcolor = dodgerblue,
     }
 
\newcommand{\uniformMmin}{$0.8$\xspace}
\newcommand{\uniformMmax}{$2.3$\xspace}
\newcommand{\gaussianMmean}{$1.33$\xspace}
\newcommand{\gaussianMstd}{$0.09$\xspace}
\begin{document}

% personalized comments
\title{Constraining the lensing of binary neutron stars from their stochastic background}

%%%%%Authors%%%%%%%%%%%%%%%%%
%[0000-0002-7387-6754]
\author{Riccardo Buscicchio}
\email{riccardo@star.sr.bham.ac.uk}
\affiliation{School of Physics \& Astronomy and Institute for Gravitational Wave Astronomy, University of Birmingham, Birmingham, B15 2TT, UK}

% [0000-0002-2527-0213]
\author{Christopher J.\ Moore}
%\email{cmoore@star.sr.bham.ac.uk}
\affiliation{School of Physics \& Astronomy and Institute for Gravitational Wave Astronomy, University of Birmingham, Birmingham, B15 2TT, UK}

% [0000-0003-4984-0775]
\author{Geraint Pratten}
%\email{gpratten@star.sr.bham.ac.uk}
\affiliation{School of Physics \& Astronomy and Institute for Gravitational Wave Astronomy, University of Birmingham, Birmingham, B15 2TT, UK}

% [0000-0003-1542-1791]
\author{Patricia Schmidt}
%\email{pschmidt@star.sr.bham.ac.uk}
\affiliation{School of Physics \& Astronomy and Institute for Gravitational Wave Astronomy, University of Birmingham, Birmingham, B15 2TT, UK}

% [0000-0002-6254-1617]
\author{Alberto Vecchio}
%\email{av@star.sr.bham.ac.uk}
\affiliation{School of Physics \& Astronomy and Institute for Gravitational Wave Astronomy, University of Birmingham, Birmingham, B15 2TT, UK}
%%%%%%%%%%%%%%%%%%%%%%%%%%%

\date{\today}

%%%%%%Abstract%%%%%%%%%%%%%
\begin{abstract}

Gravitational wave (GW) transients from binary neutron star (BNS) coalescences can, in principle, be subject to gravitational lensing thereby increasing the amplitude and signal-to-noise ratio. We estimate the rate of lensed BNS events resolvable by LIGO and Virgo and find that it is constrained by the current non-detection of a stochastic GW background. Following closely the formalism we developed previously~\cite{2020arXiv200604516B} in the context of binary black hole lensing, we show that at current sensitivities the fraction of BNS coalescences with lensing magnifications $\mu> 1.02$ is less than $\sim 7\times 10^{-8}$ and therefore such events should not be expected in the near future. We also make predictions for projected future sensitivities.

\end{abstract}
%%%%%%%%%%%%%%%%%%%%%%%%%%

\maketitle
% \tableofcontents{}
%
\section{\label{sec:Introduction}Introduction}
%
%
% \noindent{\bf \em Introduction~--~}
%
%
Two binary binary neutron star (BNS) mergers have been detected so far~\cite{2017PhRvL.119p1101A, 2020ApJ...892L...3A}.
The event GW190814~\cite{2020ApJ...896L..44A} may also contain a neutron star.
Forthcoming detector upgrades will provide better sensitivity, allowing us to probe ever larger spacetime volumes and detect more events~\cite{2018LRR....21....3A}.
Currently, the observed GW events involving neutron stars are loud and individually resolvable~\cite{1993PhRvD..47.2198F, 1994PhRvD..49.2658C}. 
However, many more events will lie below the threshold for detection, individually indistinguishable from the instrument noise.
All these events are drawn from the same overall population. 
The unresolvable GW events, including those involving neutron stars, will pile up across the detector bandwidth and give a stochastic background of gravitational waves (SGWB). Such signal is subject to dedicated searches by current ground-based interferometers~\cite{2019PhRvD.100f1101A, 2018PhRvL.120i1101A, 2019PhRvD.100d3023H}.

A fraction of BNS events will be gravitationally lensed; this has the effect of increasing the GW amplitude by a factor $\sqrt{\mu}$, where $\mu$ is the lensing magnification. 
The lensing of a GW depends on the intervening gravitational potential, and different events will experience different lensing magnifications.
Multiply imaged GWs, phasing and wave-optics effects may also occur depending on the specific potential~\cite{2010ARA&A..48...87T,2017PhRvD..95d4011D,2018MNRAS.476.2220L,2019ApJ...874L...2H, 2020arXiv200712709D, 2020MNRAS.495.3740P}.

In~\cite{2020arXiv200604516B} we considered the analogous situation for binary black hole (BBH) GW events. 
That paper described in detail the formalism to quantify the impact of lensing on the amplitude and detection rate of individual events, as well as the amplitude of the associated SGWB (see also~\cite{2020arXiv200603064M}).
Subsequently, we leveraged the non detection of a stochastic background to get constraints on the probability of individual BBHs being lensed.
In order to do so, we framed in a single statistical picture the observational data from GW detectors (i.e. individual events, stochastic background), their inferred properties, and the implication on the observation of lensed events.

In this paper we reapply the techniques from~\cite{2020arXiv200604516B} to the analysis of BNS events.
Using constraints on the BNS merger rate density after the first two observing runs~\cite{2019PhRvX...9c1040A}, and the confirmed non-detection of a stochastic background~\cite{2019PhRvD.100f1101A}, we report the implication on the expected number of lensed BNS observations, both in the weak and strong lensing regime.
Throughout we follow the conventions of ~\cite{2020arXiv200604516B}.

%%%%%%%%%%%%%End Introduction%%%%%%%%%%%%%%%%%%%%%%%%%%%
%%%%%%%%%%%%%%%%%%%%%%%%%%%%%%%%%%%%%%%%%%%%%%%%%%%%
%%%%%%%%%%%%%Models%%%%%%%%%%%%%%%%%%%%%%%%%%
\section{\label{sec:Models}Models}
%
%
%
% \noindent{\bf \em Models~--~}
%
%
We employ a semi-analytical model for the lensing probability $\mathrm{d}P(\mu\mid z)/\mathrm{d}\log\mu$ as a function of redshift out to redshifts $z\leq 20$.
This model applies to magnifications $\mu$ up to $\mu\leq 200$, as described in~\cite{2017PhRvD..95d4011D} (i.e. it includes both strong and weak lensing).
For details of our implementation of this lensing model we refer the interested reader to Appendix A of~\cite{2020arXiv200604516B}.

It is also necessary to model the BNS population. 
We neglect neutron star spins and matter effects (e.g.\ tides) as well as any orbital eccentricity in the binary. Under these simplifying assumptions a binary is described by the two component masses, $m_1$ and $m_2$.
We model the distribution of component masses, $p(m_1,m_2)$, in three different scenarios. The first two match those employed in the rates analysis of GWTC-1~\cite{2019PhRvX...9c1040A} (see Section VII.C). The third is included to investigate the effect of the width of the mass distribution. 
\begin{itemize}
    \item``Uniform''; the component masses $m_1, m_2$ are drawn independently from a uniform distribution in the range [$\uniformMmin M_\odot,\uniformMmax M_\odot$].
    \item ``Gaussian''; the component masses are drawn from a Gaussian distribution with mean $\gaussianMmean M_\odot$ and standard deviation $\gaussianMstd M_\odot$.
    \item ``Fixed''; all neutron star masses are equal to $1.4 M_\odot$.
\end{itemize}

We choose to model the redshift evolution of the BNS merger rate $\mathcal{R}(z)$ by tracking the star formation rate.
For details, see Equation 5 and Figure 1 of~\cite{2020arXiv200604516B}. 
In a slight deviation from the previous study, we keep the population extinction (i.e. $\lambda-\gamma$) fixed at high redshifts (i.e.\ $z\gg z_\mathrm{P}$) (following Madau-Dickinson~\cite{2014ARA&A..52..415M}) while increasing the slope of the local merger rate ($\lambda$).
We fix the local rate ($R_0$) to the estimates provided in~\cite{2019PhRvX...9c1040A}, while varying $\lambda$ (see Table~\ref{tab:models}).
We focus on results from one pipeline search only (pyCBC~\cite{2016CQGra..33u5004U}); analogous results have been computed for other pipelines (e.g.\ GstLAL~\cite{2019arXiv190108580S}) and differences are at the level of $1$\%.

\begin{table}[t]
\centering
\begin{tabular}{l | c | c | c }
\toprule
 Mass & $R_{0} \, (\text{Gpc}^{-3}\text{yr}^{-1})$ & Sensitivity & $\lambda$\\ \midrule
\multirow{2}{*}{Uniform} & \multirow{2}{*}{$800^{+1970}_{-680}$} & $\mathrm{O1+O2}$ & $5.83^{+1.54}_{-1.06}$\\
& & Design &  $3.26^{+1.76}_{-1.41}$\\ 
\midrule
\multirow{2}{*}{Gaussian} & \multirow{2}{*}{$1210^{+3230}_{-1040}$} & $\mathrm{O1+O2}$ & $5.686^{+1.6}_{-1.12}$ \\
& & Design & $3.078^{+1.85}_{-1.54}$ \\ 
\midrule
\multirow{2}{*}{Fixed} & \multirow{2}{*}{$1210^{+3230}_{-1040}$} & $\mathrm{O1+O2}$ & $5.685^{+1.6}_{-1.12}$\\
& & Design &  $3.077^{+1.85}_{-1.54}$\\ 
\bottomrule
\end{tabular}

\caption{\label{tab:models}
Parameters modeling the merger rate density.
The local rate $R_0$ for the ``Uniform'' and ``Gaussian'' cases are taken from the analysis in~\cite{2019PhRvX...9c1040A}.
For the ``Fixed'' case we use the same values as for the ``Gaussian'' case.
As described in the text, we increase the slope of the local merger rate $\lambda$ to a value that gives a SGWB signal with a marginally detectable signal-to-noise ratio of 2 at the sensitivities considered. 
}
\end{table}

By changing the value of $\lambda$, we effectively set the level of the BNS stochastic background (see Equation 7 in~\cite{2020arXiv200604516B}), including the effect of lensing, to a marginally detectable level.
As in~\cite{2020arXiv200604516B}, this is done for two different detector network sensitivities (namely the existing ``O1+O2'' sensitivity and a projected future ``Design'' sensitivity). 

This section has briefly described the various modelling assumptions made for (i) the lensing probability, (ii) the properties of the BNS population, and (iii) the cosmological evolution of the BNS merger rate. We refer the reader to the Appendices in~\cite{2020arXiv200604516B} for a more detailed discussion of the impact of these assumptions on the results of our analysis.

\begin{figure*}[ht]
\includegraphics[width=\textwidth]{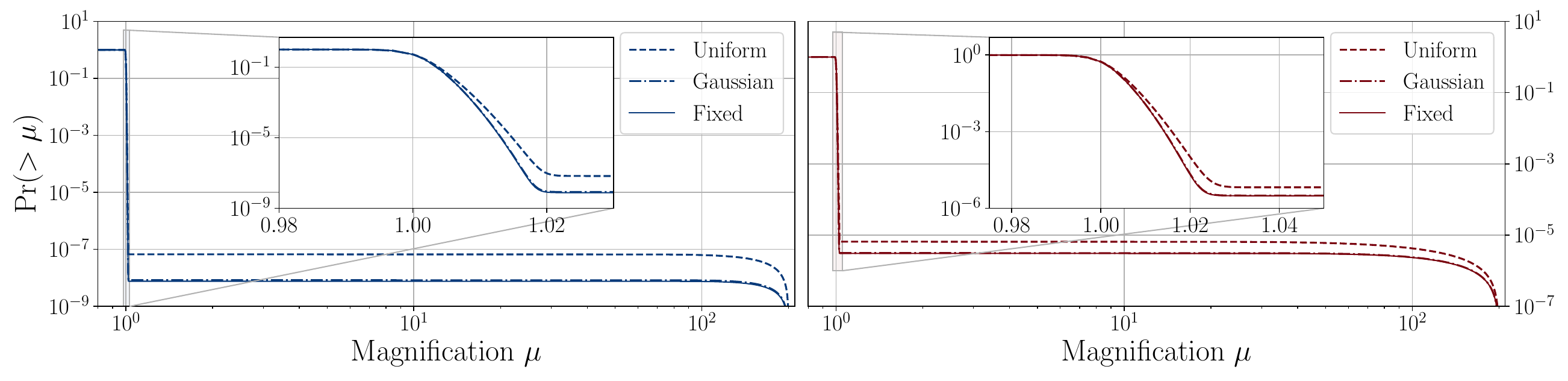}
\caption{\label{fig:lensing_fraction}
The complementary cumulative  distribution  for  the lensing probability of BNS mergers. This is constrained by the current (left, blue) or future (right, red) non-detection of a SGWB. 
The fraction of lensed transients with $\mu>1.02$ is less than $\sim 7\times 10^{-8}$ after O1 and O2; a non detection of a SGWB after 2 years of operation at design sensitivity would increase the fraction of lensed event to $\sim 7\times 10^{-6}$.
The inset plots zoom in on a narrow range of magnification around $\mu\sim 1$ where the lensing probability steeply decreases.
In our lensing model, there are  no BNS events lensed with $\mu>200$.
}
\end{figure*}

%%%%%%%%%%%%%Results%%%%%%%%%%%%%%%%%%%%%%%%%%
\section{\label{sec:Results} Results}
\begin{figure*}[ht]
\includegraphics[width=1.\textwidth]{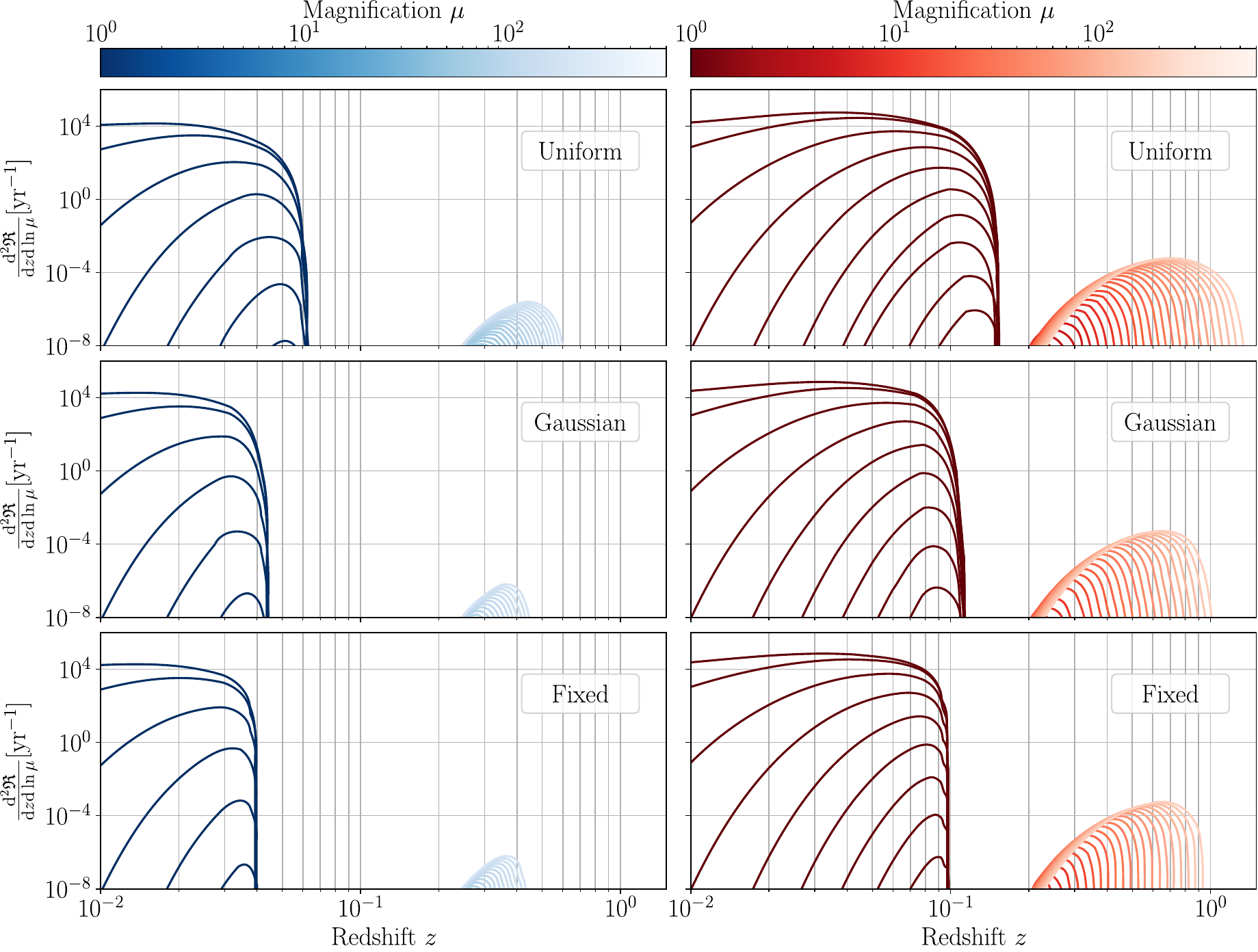}
\caption{\label{fig:lensed-rate}
Differential rate of lensed events for each redshift and logarithmic magnification bin.
Left column (blue) show results for the ``O1$+$O2'' sensitivity while the right column (red) shows results for ``Design'' sensitivity.
The three rows shows the results for the different BNS mass distributions described in the text.
Solid lines is colored according to magnification.
Moderately magnified events (i.e. $\mu<1.02$) dominate the detected population. 
For these moderately magnified events, there is a clear horizon redshift beyond which unlensed sources cannot be seen (e.g. around $z\sim0.6-0.7$ in the top-left plot). 
To see more distant events we need them to be significantly magnified. 
However, there are few large lenses at low redshifts to provide this magnification. 
There is therefore a gap out to $z\sim 3$ where large lenses are plentiful and there is a secondary contribution to the observed BNS population. The secondary contribution is more significant at ``Design'' sensitivity. 
}
\end{figure*}

%
%
% \noindent{\bf \em Results~--~}
%
%
The main result of our calculation is shown in 
Fig.~\ref{fig:lensing_fraction}. 
Here we plot the probability of a BNS event having a magnification above a certain value. 
Across all three models that we considered for the BNS masses and redshift distribution, the fraction of lensed events with magnification $\mu>1.02$ is lower than $7\times 10^{-8}$ for the ``O1$+$O2'' sensitivity.
We find it extremely unlikely on statistical ground that a significantly lensed binary neutron star will be observed in the near future.
These complementary cumulative distributions shows a large drop just above $\mu=1$ and then a very prominent plateau out to magnifications of $\mu\sim 200$ before dropping to zero. This behaviour can be understood by looking at the contribution of lensed events broken down by redshift as shown in Fig.~\ref{fig:lensed-rate}. 
The main contribution to the observed BNS population (left blob in the plots of Fig.~\ref{fig:lensed-rate}) gives the high probability near $\mu\sim1$.
The gap at intermediate redshifts visible is responsible for the plateau. And finally, in our model, there are no events with $\mu>200$, and this is responsible for the final drop off.

At design sensitivity we expect a slightly higher fraction of lensed events ($7\times 10^{-6}$). This can be seen in right panel of Fig.~\ref{fig:lensing_fraction} where a non detection of a stochastic signal yields a higher value for the plateau.

%%%%%%%%%%%%%%%%%%%%%%%%%%%%%%%%%%%%%%%
%%%%%%%%%%%%%Conclusion%%%%%%%%%%%%%%%%%%%%%%%%%%
\section{\label{sec:Discussion}Discussion}
A SGWB from BNS events has not yet been observed and is expected to be subdominant with respect to the background from BBH events.
The current non-detection places a constraint on the redshift evolution of the BNS merger rate; in particular it limits the rise in the rate at redshifts around $z\sim 2$. 
This in turn has important implications for the lensing of individual BNS events.
Here, we have used the current non-detection of a SGWB to constrain the probability that an individual BNS event is magnified by more than a certain amount.
In particular we find that the probability that $\mu>1.02$ is less than $\sim 7\times 10^{-8}$.
This probability increases slightly for detectors upgraded towards design sensitivity, but remains small. Therefore significantly lensed BNS events should not be expected in the near future.

%%%%%%%%%%%%%%%%%%%%%%%%%%%%%%%%%%%%%%%
%%%%%%%%%%%%%Acknowledgement%%%%%%%%%%%%%%%%%%%%%%%%%%
\acknowledgments
%
%
% \noindent{\bf \em Acknowledgments~--~}
%
%
The authors thanks Matteo Bianconi for useful comments. AV acknowledges the support of the Royal Society and Wolfson Foundation. PS acknowledges NWO Veni Grant No.\! 680-47-460.
Computational work was performed using the University of Birmingham's BlueBEAR HPC service.
\vfill

%%%% DO NOT CHANGE THIS %%%%
\bibliographystyle{apsrev4-2}% do not change this
\bibliography{biblio}% Produces the bibliography via BibTeX.
%%%% DO NOT CHANGE THIS %%%%

\newpage

\end{document}